\documentclass[%
 reprint,
superscriptaddress,
 amsmath,amssymb,
 aps,
floatfix,twocolumn
]{revtex4}

\usepackage{graphicx}
\graphicspath{{images/}}
\usepackage{color}
\usepackage{dcolumn}
\usepackage{bm}
\usepackage{cancel}
\usepackage{amsthm}
\theoremstyle{definition}

\usepackage{amsmath} 

\usepackage{soul}

\usepackage[ruled, linesnumbered]{algorithm2e}

\usepackage{xcolor}

\usepackage{mathtools}
\usepackage{physics} 
\DeclarePairedDelimiter\ceil{\lceil}{\rceil}

\usepackage{siunitx}

\begin{document}

\title{Mitigating errors in logical qubits}

\author{Samuel C. Smith}
 \affiliation{Centre for Engineered Quantum Systems, School of Physics, University of Sydney, Sydney, NSW 2006, Australia}

\author{Benjamin J. Brown}%
 \affiliation{IBM Quantum, T. J. Watson Research Center, Yorktown Heights, New York 10598, USA}

\affiliation{IBM Denmark, Sundkrogsgade 11, 2100 Copenhagen, Denmark}

\author{Stephen D. Bartlett}%

 \affiliation{Centre for Engineered Quantum Systems, School of Physics, University of Sydney, Sydney, NSW 2006, Australia}

\date{\today}

\begin{abstract}

Quantum error correcting codes protect quantum information, allowing for large quantum computations provided that physical error rates are sufficiently low. We combine post-selection with surface code error correction through the use of a parameterized family of \textit{exclusive decoders}, which are able to abort on decoding instances that are deemed too difficult. We develop new numerical sampling methods to quantify logical failure rates with exclusive decoders as well as the trade-off in terms of the amount of post-selection required. For the most discriminating of exclusive decoders, we demonstrate a threshold of 50\% under depolarizing noise for the surface code (or $32(1)\%$ for the fault-tolerant case with phenomenological measurement errors), and up to a quadratic improvement in logical failure rates below threshold. 
Furthermore, surprisingly, with a modest exclusion criterion, we identify a regime at low error rates where the exclusion rate decays with code distance, providing a pathway for scalable and time-efficient quantum computing with post-selection. We apply our exclusive decoder to the 15-to-1 magic state distillation protocol, and report a 75\% reduction in the number of physical qubits required, and a 60\% reduction in the total spacetime volume required, including accounting for repetitions required for post-selection.  We also consider other applications, as an error mitigation technique, and in concatenated schemes.  Our work highlights the importance of post-selection as a powerful tool in quantum error correction.

\end{abstract}

\maketitle

\section{Introduction} 

Quantum error correction (QEC) can facilitate large-scale quantum computations even with relatively noisy components.  Strategies for practical quantum error correction should have good performance in terms of both the threshold and the rate that logical failure rates decay below threshold, as well as modest hardware overheads such as the number of physical qubits per logical qubit and the complexity of the circuits used to extract the error syndrome.  

Correcting errors in a quantum code is a complex process with significant resource overheads, and some strategies for practical quantum computing make use of the much easier task of \emph{detecting} errors, post-selecting the cases where no errors have occurred \cite{Knill2005}. Post-selection has played a key role in a number of recent experimental demonstrations of quantum error correction~\cite{Harper2019, Chen2022, Postler2022, Sundaresan2023, Ye2023, Bluvenstein2023, Gupta2024, dasilva2024demonstration, hetenyi2024creating}.
Unfortunately, many direct applications of quantum error detection and post-selection are not scalable, as the probability of post-selection in a large circuit becomes vanishingly small.  Scalable approaches typically require small, fixed-distance codes to ensure a reasonable overall probability of success.  

Here, we investigate how to use the rich information contained in the error syndrome itself to make better decisions on how to post-select results, essentially interpolating between the extreme cases of error correction and error detection.  Such an approach has been considered for preparing magic states for state injection, since the circuits are so small that it is more efficient to restart the procedure when errors are detected, in order to attain higher quality final states or to reduce the overheads~\cite{li2015magic, Chamberland2019faulttolerantmagic, Gidney2021, Singh2022, Bombin2024}.  
The numerical simulations required to study this regime are challenging, and existing tools for QEC simulations such as the splitting method~\cite{Bravyi2013, Beverland2019} are not directly applicable. As a result, little is known about the impact of such post-selection methods on the performance of codes of interest such as the surface code, including thresholds and logical failure probabilities, compared with the deterministic case. 

\begin{figure} 
    \includegraphics[width=\columnwidth]{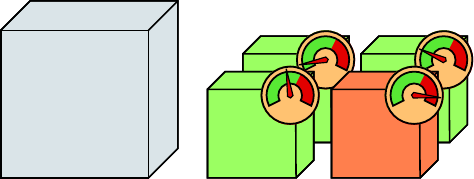} 

    \caption{\label{fig:mainidea} (left) A monolithic attempt at a medium-sized error-corrected computation, that is guaranteed to return a result but requires large overheads to reach the desired target accuracy. (right) A collection of lightweight, error-corrected attempts at the same computation. Each attempt is run at some tolerance, and if the uncertainty associated with any correction exceeds the tolerance, then the attempt is aborted. This approach is not guaranteed to return a result but failure is heralded, and when a result is returned the accuracy can be very high with low physical overheads.}

\end{figure}

We introduce an \textit{exclusive decoder} that aborts for syndromes for which the uncertainty in decoding exceeds some tolerance. We develop new numerical methods to characterise the thresholds and sub-threshold behaviour of the surface code with the exclusive decoder, demonstrating that exclusive decoding can give small surface codes the logical performance of a much larger code with a fraction of the overhead.

A common (and well-warranted) criticism of many uses of post-selection in quantum error correction and fault-tolerance is that they are not scalable.  We numerically investigate the abort probability of our exclusive decoder and find---remarkably---that the abort probability of the exclusive decoder also exhibits threshold behaviour, for all but the most intolerant instance of the decoder.  That is, there is a threshold error probability, below which the abort probability decreases exponentially in the distance of the surface code.   This behaviour suggest that an exclusive decoding strategy can find application in large-scale fault-tolerant quantum computing more broadly.

We investigate the use of exclusive decoding for medium-sized logical circuits, especially relevant on near-term devices with limited logical qubits, and for state preparation and distillation. The main idea is to break up a medium-sized error-corrected computation into many light-weight attempts at the computation, which may or may not succeed but for which failure can be heralded; see Fig.~\ref{fig:mainidea}. The result is that the large-scale error-corrected computations can be made accessible on smaller quantum devices and with lower spacetime overheads. While the computation may need to be repeated for success, much like other error mitigation strategies, the favorable behavior of the abort probabilities for the exclusive decoder suggests that such an approach will be useful in near-term implementations of fault-tolerant logic.  We also explore applications of this decoding strategy to concatenated codes, and in quantum communication.

\section{Results} 

\subsection{Exclusive decoding} 

Our exclusive decoder generalizes the concept of a decoder as used in quantum error correction.  A standard decoder is designed to take syndrome measurement data and output a logical correction, and its performance is captured by the accuracy of this correction.  However the full set of syndrome data is a large, rich data set that contains more information than just the most likely logical Pauli correction~\cite{spitz2018adaptive, Wagner2022paulichannelscanbe}, and specifically it includes information about the likelihood of correct decoding~\cite{gidney2023yoked}.  An exclusive decoder takes this same syndrome measurement data, and outputs either (1) a recommended logical correction; or (2) an abort mode where no logical correction can be identified as being correct to a sufficient degree of certainty. The degree of certainty required is a design choice that will depend on the use-case, and is a controllable parameter in the exclusive decoder. 

Concretely, the system we study here is a generalisation of the minimum-weight perfect matching (MWPM) decoder~\cite{dennis2002topological, kolmogorov2009blossom}, and we study its performance over the rotated surface code with boundaries, which allows a distance-$d$ code to be constructed out of $n = d^2$ physical qubits. The standard MWPM decoder infers errors by matching defects into pairs, with a minimum-weight matching corresponding to a most-likely error. However, on a surface code with boundaries, it is possible to modify the MWPM decoder so that it returns a minimum-weight correction for each logical equivalence class (see Appendix~\ref{app:implementation} for implementation details). With this modification, our exclusive decoder finds a global minimum-weight correction $C$, as well as alternate corrections from each other equivalence class $C \bar{L}$, where $\bar{L}$ is an $X$-type or $Z$-type logical operator. Let $\Delta$ be the unsigned weight difference between the global least-weight correction $C$ and the next-least-weight alternate correction, $C'$ that is logically inequivalent to $C$, i.e. such that $C'C$ is a logical operator. Then, the exclusive MWPM will abort if
\begin{equation} 
    1 - \frac{\Delta}{d} > c 
\end{equation} 
where $0 < c \le 1$ is a tuneable parameter that we call the \emph{exclusive tolerance}. We also define the $c = 0$ case, which is a decoder that aborts if there is any non-trivial syndrome data. With this definition, we have that a small value of $c$ means that $C$ and $C'$ must have a very large weight difference, whereas an exclusive decoder with $c=1$ is the standard MWPM decoder that will attempt to decode all syndrome errors, independent of their success probability. While the exclusive MWPM decoder assumes perfect measurements as presented here, it can be generalised to a fault-tolerant version (see Sec.~\ref{sec:FT}). 

The construction we have presented is similar to post-selection rules that have been studied previously~\cite{Bombin2024, gidney2023yoked}, and is motivated by the fact that alternate corrections that are close together in weight have a similar likelihood of occurring, making it impossible to be confident which correction is the right one. In these situations, the exclusive tolerance is used to tune how much uncertainty the decoder will tolerate before it decides to abort. Since the abort condition for the exclusive decoder depends on $\Delta$ as a fraction of the code distance, we expect the choice of tolerance to have an effect on asymptotic quantities, such as thresholds and asymptotic overheads. In the following sections, we use sophisticated numerical tools (see Appendix~\ref{app:numerical_methods}) to study the exclusive decoder as we take large code distances. 

\subsection{Exclusive decoder performance \label{sec:comparative_decoder_performance}} 

\begin{figure*}
    \centering 
    \includegraphics{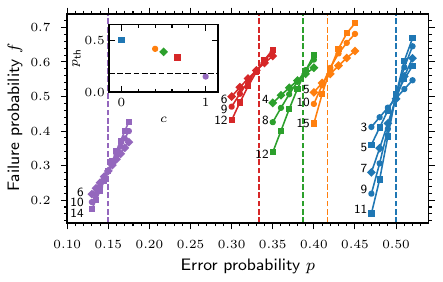} 
    \includegraphics{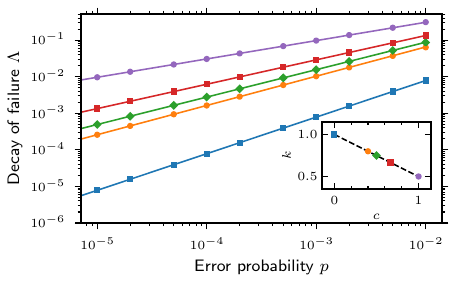} 
    \raisebox{45pt}{ \includegraphics{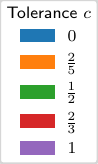}} 
    \caption{\label{fig:failure} 
    Probability of logical failure for the exclusive MWPM decoder used with the rotated surface code, against depolarizing noise and in the code-capacity setting. Logical failures are computed only on post-selected error configurations. The $c=0$ decoder is the zero-tolerance decoder that aborts on non-trivial syndrome data, and the $c=1$ decoder is the standard MWPM decoder with no post-selection. (left) Markers are data points computed using the splitting method. Solid lines are failure probabilities extracted from a critical exponents ansatz, and dashed vertical lines are the fitted threshold values. For values of the tolerance $c = 0,\ 2/5,\ 1/2,\ 2/3$, and $1$, we find threshold values $p_\text{th} = 50\%,\ 42\%,\ 38\%,\ 33\%$, and $15\%$, respectively. (left inset) Threshold values plotted against the exclusive tolerance. The horizontal dashed line is the threshold derived from the zero-rate Hashing bound. (right) Decay of logical failure probabilities below threshold. For each error probability below threshold, we fit the very general ansatz $\log(f) = d \log \Lambda(p) + \log C(p)$. The decay constants $\Lambda(p)$ are plotted with markers for each error probability. Solid lines correspond to the further ansatz Eq.~\eqref{eq:f}, which can be rewritten $\log\Lambda(p) = k \log p + \log A $. This ansatz is applicable far below threshold and only counts contributions to the failure probabilities from least-weight failing errors. The parameters $k$ and $A$ are determined by fitting, and in (right inset) we plot the fitted values $k$ against the tolerance $c$. The dashed line is Eq.~\eqref{eq:k}. We have that $k > 1/2$ for the exclusive decoder for all $c < 1$, whereas standard error correction techniques can achieve at best $k=1/2$. }
\end{figure*} 

Using simulations, we demonstrate that our exclusive decoder can lead to logical thresholds that are substantially higher than those of a standard decoder.  Equally striking, we demonstrate that the scaling behaviour of the logical failure probability below threshold can also be significantly enhanced by using an exclusive decoder.  Together, these results indicate that post-selection can provide significant additional power to error correction. 

\paragraph{Thresholds.} We estimate the threshold for the rotated surface code with the exclusive decoder under a depolarizing noise channel in the code-capacity setting (meaning perfect syndrome measurements). Logical failure probabilities are found by using the splitting method to extrapolate down from logical failure probabilities determined analytically at $p = 75\%$; see Appendix~\ref{app:numerical_methods} for details. Our results are summarised in Fig.~\ref{fig:failure}. The threshold value is $p_\text{th} = 50\%$ at zero tolerance, and decreases smoothly down to the standard MWPM decoder value of $p_\text{th} = 15\%$ as the tolerance increases to one. We remark that the depolarizing channel is known to have zero capacity for depolarizing strength in excess of $p = 25\%$, due to the no-cloning theorem~\cite{bruss1998optimal, Cerf2000, cubitt2008structure}. The exclusive decoder is able to exceed this bound due to the use of post-selection, and we discuss the exclusive decoder further in the context of quantum channel capacities in Sec.~\ref{sec:discussion}. 

One can gain some intuition about this improved decoder performance as follows. Errors that cause failure with the exclusive decoder are typically much less likely than errors that cause failure with a standard decoder, as will be discussed in Sec.~\ref{sec:comparative_decoder_performance}. Additionally, there are combinatoric factors that attach to standard decoders that count the number of ways that the $d/2$ Pauli operators that make up a least-weight failing error can be placed onto the support of a least-weight logical operator. This entropic contribution is not present for a zero-tolerance exclusive decoder, and is small (relative to standard decoding) for low-tolerance exclusive decoders. This means that both probabilistic considerations and entropic considerations are more favourable towards the exclusive decoder. Large error probabilities are therefore required before entropic factors will favour a failing error configuration, and this leads to a large threshold. 

\paragraph{Sub-threshold scaling of logical failure probabilities.} Along with improved thresholds, our exclusive decoder provides up to a quadratic reduction in the observed logical failure probability for error probabilities below threshold. Since this suppression is more rapid than can be achieved with standard error correction techniques on the surface code, we say that errors are super-suppressed below threshold. Scaling improvements of this type can be used to reduce resource overheads, since fewer physical qubits are required to achieve commensurate logical failure probabilities compared to the standard approach. 

The failure probabilities of quantum error correcting codes well below threshold are determined by the weight and number of least-weight failing errors. In particular, if the weight of the least-weight failing error is some fraction $k$ of the code distance, then an ansatz logical failure probabilities below threshold can be written
\begin{equation} 
    f = C (A p)^{kd} \label{eq:f} 
\end{equation}
where $A^{kd}$ counts the number of least-weight failing errors. Since we expect to operate quantum error correction well below threshold, this ansatz is the key relation that determines the resource overheads for fault-tolerant quantum computation. In particular, the least fractional weight $k$ is a key quantity of interest. 

For standard surface code error correction, there is no decoder that can correct all errors of weight up to and including $w = \ceil{d/2}$, which sets an upper bound $k \leq 1/2$. To see this, take a logical operator $\bar{L}$ of weight $d$. We can split $\bar{L}$ into two parts $\bar{L} = E + E'$ with weights $w = \ceil{d/2}$ and $w' = \lfloor(d - 1)/2\rfloor$. Since $E$ and $E'$ have the same syndrome, it is impossible to simultaneously correct both of them. A signature of this problematic error configuration is that the two alternate corrections are very close in weight; in the example above the weight difference between the alternate corrections was equal to one. The exclusive decoder safeguards against this by aborting if there exists minimum weight corrections in alternate sectors that are close together in weight.

With the exclusive decoder, let $w$ denote the weight of a failing error $E$ that does not lead to an abort. Let $w'$ denote the least-weight correction $E'$ in an alternate sector, then the failure condition implies that $w - w' \geq d(1 - c)$. Additionally, we must have $w + w' \geq d$, since $E + E'$ gives a logical operator. This gives a lower bound on the weight of failing errors. The exclusive decoder can abort or correct any error with weight $w < w_\text{min}$, which expressed as a fraction of the code distance is $w_\text{min} = kd$ with $k$ given 
\begin{equation} 
    k = 1 - \frac{c}{2}\,. \label{eq:k} 
\end{equation} 
We show in Fig.~\ref{fig:failure} the subthreshold failure probabilities, and in particular we extract the least fractional weight $k$ as a function of the exclusive tolerance, which shows the agreement with Eq.~\eqref{eq:k}.

\subsection{Abort probabilities} 

A key quantity of interest for an exclusive decoder is the probability with which it will abort. The behavior of the abort probability will capture the trade-off being made for reduced logical failure probabilities and boosted logical thresholds, in terms of (say) an increased time cost in a `repeat until success' approach. Our analysis shows that the abort probabilities for non-zero tolerance exhibits threshold behaviour, meaning that the probability of aborting decreases exponentially as a function of the code distance for error probabilities less than the abort threshold.  The existence of such an abort threshold suggests that the exclusive decoder may find application in scalable approaches to fault-tolerance as well as near-term uses.

\begin{figure*} 
    \includegraphics{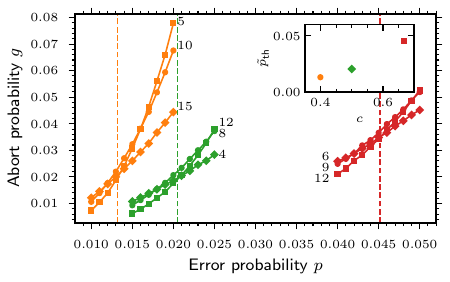} 
    \includegraphics{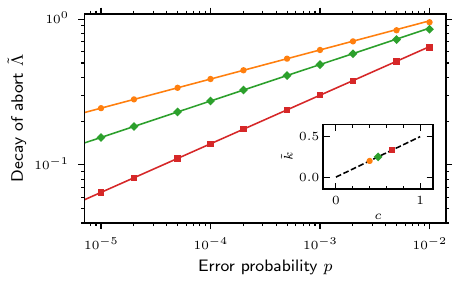}
    \raisebox{45pt}{ \includegraphics{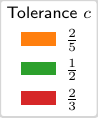}} 
    \caption{
    \label{fig:abort}
    Probability that the exclusive MWPM decoder over the rotated surface code will abort. The noise model is depolarizing noise in the code-capacity setting. (left) Markers are data points computed using direct Monte Carlo estimation, and crossings indicate the existence of a threshold with respect to abort probabilities. Solid lines are abort probabilities extracted from a critical exponents ansatz, and dashed vertical lines are the fitted threshold values. For tolerances $c = 2/5,\ 1/2,\ 2/3,$ we observe abort threshold values $\tilde{p}_\text{th} = 1.3(1)\%,\ 2.1(1)\%,\ 4.5(1)\%,$ respectively. Note that only data for tolerances in the range $0 < c < 1$ are shown. The zero-tolerance case is omitted because it does not possess a threshold for aborts, and the standard MWPM case is omitted because there are no aborts. (left inset) Threshold values for the abort probability plotted against the exclusive tolerance. (right) Decay of abort probabilities below threshold. For each error probability below threshold, we fit the very general ansatz $\log(g) = d \log \tilde{\Lambda}(p) + \log \tilde{C}(p)$. The decay constants $\tilde{\Lambda}(p)$ are plotted with markers for each error probability. Solid lines correspond to the further ansatz Eq.~\eqref{eq:ktilde}, applicable far below threshold, which can also be written $\log \tilde{\Lambda}(p) = \tilde{k} \log p + \log \tilde{A}$. The parameters $\tilde{k}$ and $\tilde{A}$ are determined by fitting, and in (right inset) we plot the fitted values of $\tilde{k}$ against the tolerance $c$. The dashed line is Eq.~\eqref{eq:ktilde}. We remark that the abort probability decays more slowly as the tolerance to uncertainty is decreased.
    } 
\end{figure*}

\paragraph{Abort thresholds.}  In Fig.~\ref{fig:abort}, we demonstrate numerical evidence of abort thresholds for non-zero tolerance $c>0$. For the case of zero-tolerance, the exclusive decoder does not have an abort threshold, and we characterize the abort probabilities for the zero-tolerance decoder in App.~\ref{app:additional_numerical_results}. 

For all tolerances $c>0$, the value of the abort threshold is much smaller than the value of the logical threshold, and also for the value of the logical threshold for standard MWPM decoding. This is as expected; the exclusive decoder will definitely abort on any low-weight error that causes standard MWPM decoding to fail, but additionally will also abort on some errors that would have been successfully corrected by standard MWPM decoding, since the exclusive decoder can be quite cautious.  This fact is significant, and has consequences for the usefulness of the exclusive decoder for large codes. There are two regimes in which we can imagine operating the exclusive decoder: (i) below the abort threshold; and (ii) above the abort threshold but still below the logical threshold. In the first regime, the exclusive decoder can be used on large codes since the time-cost does not blow up with the size of the code, and we discuss these asymptotic overheads in Sec.~\ref{sec:overheads}. In the second regime, the practical usefulness of the exclusive decoder is likely limited to small codes. We further discuss the scope of use-cases in Sec.~\ref{sec:discussion}. 

\paragraph{Scaling of abort probabilities.}  As with logical failure rates, the abort thresholds only tell part of the story, and we also investigate the scaling of the abort probability as a function of code distance both above and below the abort threshold. Below the abort threshold, we study the decay of abort probabilities asymptotically. Above the abort threshold, we develop an empirical ansatz to calculate abort probabilities for finite-size codes. 

In the regime of error probabilities below the abort threshold of the protocol, we can expect a similar ansatz to Eq.~\eqref{eq:f} to hold. That is, we expect to describe abort probabilities by 
\begin{equation} 
    g = \tilde{C} p (\tilde{A} p)^{\tilde{k} d} \,. \label{eq:g} 
\end{equation} 
Here, $\tilde{k} d$ is one less than the weight of the least-weight error that will lead to an abort, which gives rise to an additional factor $p$ as compared to Eq.~\eqref{eq:f}. The factor $\tilde{A}^{\tilde{k} d}$ counts the number of least-weight aborting errors. To find $\tilde{k}$, take an error $E$ of weight $w \leq d/2$ that leads to an abort, and let $w'$ denote the weight of the least-weight correction $E'$ that is inequivalent to $E$. We have $w' - w < (1 - c)d$ and $w + w' \geq d$ holding simultaneously, which gives $2w > cd$, or equivalently
\begin{equation} 
    \tilde{k} = \frac{c}{2} \,. \label{eq:ktilde}
\end{equation}  
We remark that the more exclusive the decoder is, in terms of being intolerant of uncertainty, the more slowly the abort probability decays in the code distance. In Fig.~\ref{fig:abort}, we show the scaling of the abort probabilities with the code distance, and in particular we see good agreement with Eq.~\eqref{eq:ktilde}. 

\begin{figure} 
    \includegraphics{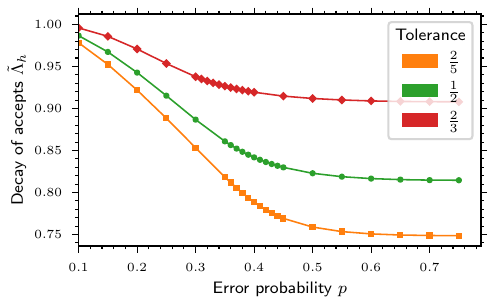} 
    \caption{
    \label{fig:abort_above_threshold}
        The decay of acceptance probabilities as a function of the number of qubits in the code, above the abort threshold. Data is for the exclusive decoder with depolarizing noise on the rotated surface code, in the code-capacity setting. For each error probability above the abort threshold, we fit the ansatz Eq.~\eqref{eq:h}, which can also be written $\log h = n \log \tilde{\Lambda}_h(p) + \log \tilde{C}_h(p)$, where $n$ is the number of qubits and in this case we have $n = d^2$. Details of these fits can be found in Appendix~\ref{app:supporting_figures}. We extract and plot $\tilde{\Lambda}_h$ as a function of the error probability and for each value of the tolerance.
    }
\end{figure}

We are also interested in the abort probabilities for finite-size codes for error probabilities that are larger than the abort threshold, but smaller than the logical threshold. It is less clear a priori how abort probabilities will behave above the abort threshold. However, we empirically observe that the acceptance probability, defined $h = 1 - g$, decays exponentially in the number of qubits $n$ in the code. That is, we have 
\begin{equation} 
    h = \tilde{C}_h(p) \tilde{\Lambda}_h(p)^n \label{eq:h} 
\end{equation} 
where $\tilde{C}_h$ can be interpreted as a finite-size offset, and $ \tilde{\Lambda}_h < 1$ is the decay factor, and $n$ is the number of qubits with $n = d^2$ for the rotated surface code. In Fig.~\ref{fig:abort_above_threshold}, we extract decay factor $\tilde{\Lambda}_h$ for a range of error probabilities and exclusive tolerances. Since above the abort threshold, the practical use of the exclusive decoder is limited to finite-size codes, Eq.~\eqref{eq:h} can be useful to apply to small code instances.

\subsection{Fault tolerant syndrome measurements \label{sec:FT}} 

We now turn to the question of fault-tolerance and imperfect measurements.  The general idea of an exclusive decoder maps straightforwardly into the fault-tolerant case, and in particular the MWPM-based decoders used for the surface code are naturally made fault-tolerant~\cite{dennis2002topological}.

There are a few subtleties that offer challenges for characterising the fault-tolerant case numerically.  A common setting in which decoders are numerically benchmarked in the fault-tolerant setting involves taking periodic time-like and space-like boundary conditions. A minor difficulty arises when generalizing exclusive MWPM to this setting, since it involves fixing the parity of matches over an arbitrary subset of edges in the decoding graph. This should be compared to the code-capacity regime with rough and smooth boundaries, where we only needed to fix the parity of matches between bulk and boundary vertices (see Appendix \ref{app:implementation}). This problem is still solvable in the fault-tolerant setting, since the decoder studied in Ref.~\cite{Sahay2022} utilizes a heuristic solution to this problem that can find low-weight corrections in alternate sectors, and this subroutine can be used to implement a fault-tolerant version of exclusive MWPM. Then, the abort condition can treat time-like errors in exactly the same way as space-like errors.

We do study the exclusive decoder at zero-tolerance and with imperfect measurements in Appendix \ref{app:additional_numerical_results}. There, we set faulty measurements to occur with probability $p_m = 2 p / 3$, and we report a fault-tolerant threshold value $p = 32(1)\%$. As well as this, we develop a sophisticated ansatz that captures subthreshold failure probabilities, and observe the scaling relationship $f \propto p^d$, as expected. Additionally, we introduce in Appendix \ref{app:union-find} the \textit{exclusive union-find decoder}, which also generalizes to the fault-tolerant regime straightforwardly. We expect Eq.~\eqref{eq:k} and Eq.~\eqref{eq:ktilde} to still hold for the fault-tolerant exclusive MWPM decoder, and the fault-tolerant exclusive union-find decoder.  

Finally, we remark that future work is still required to generalize homology-based decoders, such as maximum-likelihood decoding or exclusive decoding, to the fault-tolerant setting with open time-like boundaries in a sliding-windows fashion~\cite{skoric2023parallel, Tan2023}.

\section{Discussion \label{sec:discussion}}

The post-selection scheme presented here is a promising approach to protecting against noise in the nascent era of fault-tolerance, where logical qubits may be limited in size. Post-selection can enable more reliable calculations out of early-stage logical qubits that may not be sufficient for a particular use-case using standard decoders.  They can also allow for a reduction in resource overheads.  We now explore these applications in some detail.

\subsection{Resource overheads \label{sec:overheads}} 

The exclusive decoder filters out the noisiest shots in a quantum computation. In the regime of limited logical qubits, this decoder allows access to deeper logical circuits, and can reduce footprints for fault-tolerant primitives.  We view this regime of circuits as a natural sequel to NISQ (noisy intermediate-scale quantum) circuits, where logical encodings are combined with post-selection to access early-stage fault tolerance on noisy intermediate-scale devices. 

\paragraph{Deeper logical circuits.} We analyze this trade-off in the low-error-probability regime. A circuit volume $q$ can be executed using logical qubits at failure probability $f$, up to a target accuracy $\epsilon = qf$. In the non-exclusive case, the scaling of $f$ in the code distance is at best $f = p^{d/2}$. In the exclusive case, the scaling is improved to $f = p^{k d}$ for some $1/2 \leq k \leq 1$. However, this comes at the cost of repetitions that must be performed 
\begin{equation} 
    R = (1 - p^{\tilde{k} d})^{- q} \label{eq:R}
\end{equation} 
where $\tilde{k} = 1 - k$, since the computation must be restarted if any error correction cycle returns an abort mode. 

If we take a fixed number of physical qubits, then using surface codes of a certain distance will set an accessible circuit volume $q_0 = \epsilon p^{-d/2}$. This can be increased to $q = \epsilon p^{-kd}$ using exclusive decoding techniques. If we only attempt to achieve a constant factor improvement, then we can set a fixed $a = \sqrt{\epsilon} q / q_0$. We also define the rescaled circuit volume $x = q_0 / \sqrt{\epsilon}$. We can rewrite the number of repetitions required by Eq.~\eqref{eq:R} in terms of $x$ and $a$, giving $R = (1 - a/x)^{-ax} \approx \exp(a^2)$. This approximation is accurate provided that $q$ is small compared to $q_0^2 / \epsilon$. Finally, we can re-arrange to express the factor increase in circuit depth achievable in terms of the number of repetitions we are willing to perform
\begin{equation} 
    q = \sqrt{\frac{\log R}{\epsilon}} q_0 \,.
\end{equation} 
For an entire algorithm where we might consider $\epsilon \approx 1$, then this trade-off allows, for instance, a 2X increase in the accessible circuit depth, in exchange for repeating the circuit approximately 55 times. We remark that this trade-off is only available up to an upper bound $q \leq q_0^2 / \epsilon$ that arises from the fact that the maximum boost to logical failure probabilities is quadratic, which is achieved by the zero-tolerance decoder. 

\paragraph{Reduced qubit overheads.} An alternative but equivalent view is that we may want to execute a fixed size circuit, but using a smaller number of physical qubits, i.e., using a smaller distance code. We start with a fixed circuit of size $q$, executed using surface codes of distance $d_0$, and with $m_0$ physical qubits. Using a exclusive decoder with tolerance $c = 1 - 2k$, this circuit can alternatively be implemented using distance $d = d_0 / (2k)$ codes, and requiring $m = m_0 / (4k)^2$ physical qubits. The target accuracy $\epsilon$ remains unchanged. 

If we assume that the number of repetitions required is not unreasonably large, so that $\log R$ is small compared to $q$, then we can rewrite Eq.~\eqref{eq:R} in terms of circuit quantities as 
\begin{equation} 
    \log R = \epsilon^{2 \sqrt{m / m_0} - 1} q^{2(1 - \sqrt{m/m_0})} \,. \label{eq:footprint} 
\end{equation} 
This equation suggests that the exclusive decoder will allow large footprint reductions, at a modest time-cost, when applied to circuits with low target accuracy $\epsilon$, but that also have a small number of gates $q$. 

\paragraph{Key application: distilling magic states.} Motivated by conclusions drawn from Eq.~\eqref{eq:footprint}, we consider the exclusive decoder applied to medium-sized state preparation circuits. The exclusive decoder is well-suited to this use-case, since  the output of a state preparation circuit is a quantum state that needs to be extremely high fidelity, since it will get consumed in a much larger circuit. This small target accuracy allows an exclusive decoder to be used for a footprint reduction, without also incurring a large overhead in terms of repetitions required. 

As an example, we do a rough analysis of the 15-to-1 magic state distillation circuit~\cite{bravyi2005universal}, in the fault-tolerant setting.  We simulate the zero-tolerance exclusive decoder using a distance $d = 4$ code in the fault-tolerant regime, and compare to a standard decoder using a distance $d = 8$ code. We assume the decoders have similar logical performance, since the weight of the least-weight failing error is equal to four in both cases. If anything, we expect this assumption to be quite generous towards the standard decoder since there are many more failure mechanisms on the distance-8 code compared to the distance-4 code. We use values for $q$ and $\epsilon$ based on the 15-to-1 construction from Ref.~\cite{Litinski2019}, that has $q \approx 90$ error correction cycles, an output accuracy limited by $\epsilon_T \approx 10p^3 \approx 10^{-11}$, due to errors that occur during faulty $T$ measurement. We set an error probability $p = 10^{-4}$. Using the detailed ansatz in Appendix~\ref{app:additional_numerical_results}, we have an expression for $f$ and $g$, and can compute the total failure probability of the circuit with the zero-tolerance decoder, $\epsilon = fq \approx 5.8 \times 10^{-14}$. This means approximately one in two hundred failures will be due to surface code failures, as opposed to faulty $T$ measurement failures, so surface code failures contribute negligibly to the total circuit accuracy with these parameters. We can also compute the amount to circuit repetitions required using Eq.~\eqref{eq:ztft_abort}  with $R = (1 - g)^{-q}$, and report that only 3.2 circuit repetitions are required on average due to the use of the zero-tolerance decoder. We compare to the case of a standard decoder using a distance $d = 8$ code with no repetitions, and assume this comparison gives equal logical performance. Making this comparison, we report that the exclusive decoder requires 25\% of the physical qubits used by the standard decoder, and also requires 40\% of the total spacetime volume, including with repetitions, in order to achieve commensurate logical performance with the standard decoder. We remark that the logical performance of the zero-tolerance decoder was better than it needed to be, since the logical accuracy of the distillation protocol is bottlenecked by faulty $T$ measurement, not by surface code failures. We could have chosen a non-zero tolerance in order to reduce the number of repetitions, at the expense of larger logical failure probabilities, provided we maintain $\epsilon \ll \epsilon_T$. In general the optimal tolerance will balance savings that result from using a small code distance, with the repetitions that are required when the decoder aborts.

We emphasize the usefulness of exclusive decoding applied to a range of medium-sized state preparation tasks, beyond magic state distillation. Especially, the exclusive decoder is well suited to quantum computing paradigms that involve breaking large computational tasks into smaller and independent state preparation tasks, such as with error-correcting teleportation~\cite{Knill2005} or quantum dynamical programming~\cite{son2024quantum}. These paradigms can be cheap in terms of the on-line part of the computation~\cite{Huggins2024, marvian2024universal}, and the off-line part of the computation can be done using low-distance codes with exclusive decoding. 

\subsection{Other applications} 

The exclusive decoder could also prove useful in schemes that use code-concatenation to achieve higher encoding rates, such as Refs.~\cite{pattison2023hierarchical, gidney2023yoked}. In this setting, we envision the abort mode of the exclusive decoder as heralding a logical erasure that is dealt with by an outer code. The exclusive decoder can then be viewed as a gadget that converts Pauli noise into erasure noise. This is similar in spirit to previous work on erasure conversion via qubit engineering~\cite{Levine2024, kubica2023erasure, gu2023faulttolerant}, or using concatenation with small inner codes~\cite{Li2023}. An exclusive-style decoder that extracts soft information about logical Pauli noise for code-concatenation was also studied in Ref.~\cite{gidney2023yoked}. The success of this approach is based on the fact that structured or biased noise can be exploited to achieve very large performance gains~\cite{bonilla2021xzzx}. The quantum erasure channel, in particular, is arguably the easiest-to-correct source of noise that can affect a quantum computation---it has non-zero capacity up to $p < 50\%$~\cite{Bennett1997}, it requires a weight-$d$ erasure error to fail a distance-$d$ code, and it admits high performing and accurate decoders \cite{delfosse2020lineartime}. The erasure channel is also conceptually straightforward relative to Pauli noise. Insights gained from first considering erasure are often found also to be useful in the more general noise setting, both in the context of designing decoders~\cite{delfosse2020lineartime, delfosse2021almostlinear, delfosse2022toward, connolly2023fast}, and designing error correcting codes~\cite{arikan2009channel, kudekar2011threshold, kudekar2016reedmuller}. The conversion of Pauli noise to erasure noise at the surface code level by the exclusive decoder may help to simplify the decoding problem faced by the outer code in a concatenated scheme. If the outer code is a more exotic kind of code, for which decoding is more challenging, then this will be an especially useful approach.  

Another use-case of the abort mode is to signal the level of sophistication of the decoder that should be used to correct the error. A common design paradigm for high-performance real-time decoders involves hierarchical decoding \cite{delfosse2020hierarchical, ueno2022neoqec, Meinerz2022, ravi2023better, das2022lilliput, Chamberland2023, Gicev2023, Paler2023, Smith2023}, and we point the reader to Ref.~\cite{battistel2023realtime} for a recent perspective. In this paradigm, the first decoding stage is fast and efficient, but may make mistakes more readily due to a lack of global syndrome information, for example. This first stage is designed to correct the majority of sparse error configurations. The second stage is powerful but costly, correcting the difficult corrections that are leftover by the first stage. The exclusive techniques explored here are capable of filtering the easy errors from the difficult errors, allowing sophisticated decoding to be called only when needed. As an example, the pre-decoder studied in Ref.~\cite{Smith2023} is fast and efficient, but can in rare circumstances grow the weight of an adversarial error by a factor $3/2$. This leads to failure at the second stage of decoding, and reduces the effective code distance. However, when combined with suitable exclusive decoding, the protocol can correct up to the full code distance, whilst also correcting typical sparse errors with high performance, maintaining the best of both worlds. 

We also consider how our exclusive decoder fits into the existing framework that describes quantum communication through noisy channels. In that framework, a channel with capacity $Q$ can be used to transmit $m$ qubits with $n$ uses of the channel, provided that $m/n \leq Q$, and with error vanishing in $n$ (see Ref.~\cite{Holevo2020, gyongyosi2018survey} for two surveys). However, the capacity depends on the presence or absence of free classical communication. A forward classical side-channel from sender to receiver will not increase the capacity~\cite{Bennett1996}. A backwards classical side-channel will increase the capacity to $Q_B \geq Q$~\cite{Leung2008}, and a two-way classical side channel will increase the capacity further to $Q_2 \geq Q_B \geq Q$~\cite{Bennett1996, Bennett1997}. A quantum error-correcting memory with post-selection can be viewed as a communication protocol with two-way classical communication via error-correcting teleportation~\cite{Knill2005, aschauer2005quantum}. In this setting, transmission of a qubit is achieved by first establishing a high-fidelity logical Bell pair shared between sender and receiver, and then a logical qubit can be teleported through the Bell pair, requiring some forward classical communication to pick a suitable Pauli correction~\cite{Bennett1996a}. Many attempts can be made at establishing this Bell pair, since the receiver can ask the sender to keep trying until an exclusive decoder does not return an abort on the receivers half of the Bell pair. In this picture, QEC techniques are directly used to protect against noisy channels. This approach complements the approach of Ref.~\cite{christandl2024faulttolerant, belzig2022faulttolerant}, where QEC techniques are used to fault-tolerantly implement encoding and decoding circuits for communication with noisy local gates. Since we have used the surface code in the current work, which has zero rate, we cannot hope to learn anything about the region $Q_B > 0$. However, it does raise an interesting question: is there a separation between $Q_B > 0$, and the region where communication is not possible. Put another way, is it possible to pass a constant number of qubits through a channel that has zero capacity? Unfortunately, the zero-tolerance surface code protocol is not an example, since the 50\% depolarizing channel with a backward side-channel is known to have positive capacity~\cite{Bennett1996}. However, the numerical tools we have developed allow us to study the asymptotics of error-correcting codes with huge amounts of post-selection, and future work could explore better-performing codes to transmit information through very noisy channels. 

\begin{acknowledgments}
We are grateful to V. Siddhu, G. Smith and M. Tomamichel for helpful conversations and comments.
BJB thanks the Center for Quantum Devices at the University of Copenhagen for their hospitality.
This work is supported by the Australian Research Council via the Centre of Excellence in Engineered Quantum Systems (EQUS) project number CE170100009, and by the ARO under Grant Number: W911NF-21-1-0007. 
\end{acknowledgments}


\appendix 

\section{Implementing exclusive MWPM decoding \label{app:implementation}}


Here we describe how to construct a matching graph so that the matching returned by a MWPM algorithm will correspond to a least-weight logical operator within a specified logical sector. We follow the construction of Ref.~\cite{Hutter2014}, and focus on matching graphs for one sector of the surface code, with boundaries. 

\begin{figure} 
\includegraphics[]{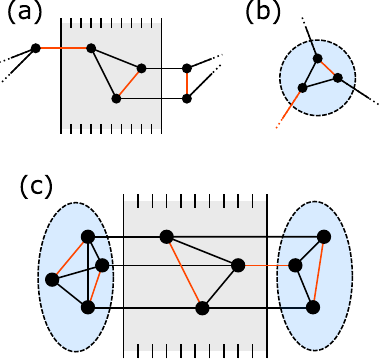}
\caption{\label{fig:matchings} (a) The matching graph for standard MWPM for defects created by $X$-type Pauli errors that condense on the smooth boundaries. There are three bulk defects, and therefore three boundary defects, with each bulk defect connected to its nearest boundary. The boundary defects are fully connected with each other, with some connections wrapping around the figure, as indicated by edges that terminate in dots. A minimum weight perfect matching is indicated in red. (b) A subset of a graph containing an odd parity of vertices. The parity of matches between the subset and its complement is fixed to be odd. (c) The exclusive matching graph has one set of boundary nodes for each boundary. On the right boundary, there is one boundary vertex for every defect in the code, which fixes an odd parity of matches to the right boundary. On the left boundary, there is one boundary vertex for every bulk defect, plus an additionally boundary vertex to fix an even parity of matches to the left boundary. }
\end{figure}

The matching graph used for standard MWPM is shown in Fig.~\ref{fig:matchings}. It has bulk vertices that correspond to defects in the code. Each pair of bulk defects is connected with an edge, with a weight that reflects the probability of an error chain leading to a creation of the defect pair. The standard matching graph also has boundary vertices that allow defects to be matched to the boundary of the code. The connectivity of the boundary vertices allows every bulk node the freedom to be matched to its nearest boundary, whilst placing no further constraint on the matching. 

We want to modify this graph in order to fix the parity of matches that connect defects to a particular boundary. Very generally, if we choose some subset of the vertices of a graph $V'$, then the parity of matches included in a perfect matching that pair vertices in $V'$ with vertices not in $V'$ is fixed by the size of $V'$. We therefore need to modify the matching graph to fix the number of virtual boundary nodes at each boundary. A matching graph that does this is shown in Fig.~\ref{fig:matchings}. 

Note the small detail that, for every bulk defect, it is now necessary to include an edge between the defect and both boundaries. This inclusion is important because, in the exclusive setting, it can sometimes be optimal to match a defect to the more distant boundary. This can be necessary in order to satisfy the parity constraints on that boundary, so that other bulk defects are free to be matched with each other in a low-weight correction. This is contrasted to the standard MWPM decoder, where it is never optimal to match a defect to the more distant boundary. 

\section{Numerical methods \label{app:numerical_methods}}

In this work, we develop new numerical methods that are used to compute the post-selected failure probabilities and acceptance probabilities presented in the main text. These methods make use of the splitting method~\cite{Bravyi2013, Beverland2019} as a subroutine, which we review briefly in subsection~\ref{ssec:the-splitting-method}. In subsection~\ref{ssec:splitting-individual-logicals} we show how the splitting method is used to develop the numerical methods that are used in this work.

\subsection{The splitting method\label{ssec:the-splitting-method}} 

The splitting method allows for the simulation of extremely rare events by computing and multiplying together the ratios of probabilities of a sequence of increasingly rare events. We give only a high-level overview of aspects of the splitting method that are relevant to this work, and we point the interested reader to Ref.~\cite{Bravyi2013} or Ref.~\cite{Beverland2019} for a more detailed explanation of this method applied to quantum error correction. 

The splitting method is very useful, for instance, for computing failure probabilities of quantum error-correcting codes at small error probabilities. In particular, denote by $\mathbb{P}_j$ a family of error models, where we may imagine large $j$ corresponding to a low-error probability regime, and denote by $\mathcal{F}$ the set of error configurations that will cause a given decoder to fail. The splitting method makes use of the acceptance ratio method from \cite{Bennett1976}, for estimating the free-energy difference between two canonical ensembles. The acceptance ratio method provides a means to compute the ratio 
\begin{equation} 
	R_j = \frac{ \mathbb{P}_{j + 1}(\mathcal{F}) }{  \mathbb{P}_{j} (\mathcal{F})}. \label{eq:splitting} 
\end{equation} 
provided samples of failing errors, i.e., samples drawn from $\mathbb{P}_j(E | \mathcal{F})$ and from $\mathbb{P}_{j+1}(E | \mathcal{F})$. Then, provided that one has access to the direct probability $\mathbb{P}_0(\mathcal{F})$, for example by direct Monte-Carlo estimation, then the probability of failure at any error probability can be computed as a product of these ratios: 
\begin{equation} 
	\mathbb{P}_j(\mathcal{F}) = \prod_{j' = 0}^{j - 1} R_{j'}\mathbb{P}_0(\mathcal{F}) 
\end{equation} 
The quantity $\mathbb{P}_0(\mathcal{F})$ may be computed using standard Monte-Carlo techniques at some error probability where it is feasible to do so. This will likely be close to threshold, where the failure probability does not vanish in the code distance, and in practice a rough estimate of the threshold can be found by a preliminary sweep. Alternatively, $\mathbb{P}_0(\mathcal{F})$ may be calculated using analytic methods where possible. 

We remark that the splitting method was used to compute abort probabilities of the exclusive decoder below the abort threshold. The method was used as described, except with the failing error set $\mathcal{F}$ replaced with the abort set $\mathcal{A}$. Direct Monte-Carlo methods were used to compute abort probabilities close to the abort threshold. More sophisticated methods were required to compute post-selected failure probabilities, as well as for acceptance probabilities above the abort threshold, and these methods are developed in the next section. 

\subsection{Splitting individual logicals\label{ssec:splitting-individual-logicals}} 

In this section, we develop a new method that is used to compute the post-selected failure probabilities and acceptance probabilities presented in the main text.

The splitting method cannot be directly applied in the current setting, for two reasons. The first reason is that post-selected failure probabilities are conditional on the decoder accepting the syndrome history. That is, if we denote by $\mathcal{A}$ the set of errors that lead to aborts, then post-selected failure probabilities are given $f = \mathbb{P}(\mathcal{F} |  \mathcal{A}^c)$. To use the splitting method, we would need to compute the ratios of conditional probabilities, and this would require being able to efficiently compute the conditional probability density function $\mathbb{P}_j(E | \mathcal{A}^c)$, which we cannot do. The second problem is that it is difficult to even sample from $P_j(E | \mathcal{A}^c)$ for low error probabilities. If only local steps are used by a sampler, the sampler is no longer ergodic because it cannot couple failing errors from different logical sectors without passing through some error that causes the decoder to abort. On the other hand, if non-local steps are used by a sampler, then a step that couples logical sectors is going to occur with a probability that is exponentially small in the code distance.

To overcome these difficulties, we introduce a method that samples from each logical sector separately, and at the end combines this sample data using a normalization procedure to give the probabilities of failure due to each logical sector. Denote by $\mathcal{L}$ the set of all error configurations $E$ that are accepted by the decoder and are corrected by a Pauli correction $C$ resulting in a logical operator $EC = \bar{L}$. We then have, for instance, that $\mathcal{F} = \mathcal{X} \cup \mathcal{Y} \cup \mathcal{Z}$ for a code that contains only one logical qubit. Applying the splitting method to each logical sector separately allows us to calculate the unconditional sector probabilities: 
\begin{equation} 
    \tilde{\mathbb{P}}_j(\mathcal{L}) = c_\mathcal{L} \mathbb{P}_j(\mathcal{L})\,, \label{eq:sector_probs} 
\end{equation}
where $c_\mathcal{L}$ are as-yet unknown constants of proportionality that do not depend on $j$. 

We now make use of our knowledge of the noise model, which in the main text was a depolarizing noise model, with $\mathbb{P}_0$ corresponding to a depolarizing parameter $p = 75\%$. At this error probability, the noise model has a symmetry, which is that all Pauli operators are equally likely to occur. This fact can be exploited to relate the constants of proportionality. In particular, it guarantees that $\mathbb{P}_0(\mathcal{L})$ is independent of $\mathcal{L}$. If we fix $\tilde{P}_0(\mathcal{L}) = 1$ without loss of generality, then inspecting Eq.~\eqref{eq:sector_probs} leads us to conclude that $c_\mathcal{L}$ must also be independent of $\mathcal{L}$, and we drop the subscript and write $c_\mathcal{L} = c$.

Finally, since post-selected failure probabilities involve ratios of unconditional sector probabilities, the constant of proportionality cancels out, and we have: 
\begin{equation} 
    f_j = \frac{\sum_{\mathcal{L} \neq \mathcal{I}} \tilde{\mathbb{P}}_j(\mathcal{L}) }{\sum_\mathcal{L}  \tilde{\mathbb{P}}_j(\mathcal{L})} \,.
\end{equation} 

Acceptance probabilities above the abort threshold can be computed up to a constant of proportionality: 
\begin{equation} 
    h_j := 1 - g_j = \frac{1}{c} \sum_\mathcal{L} \tilde{\mathbb{P}}_j(\mathcal{L}) \,,
\end{equation} 
and the constant of proportionality can be removed by fixing to direct Monte-Carlo estimates of the abort probability taken near threshold. 

We made special use of properties of the depolarizing noise channel at $p = 75\%$. We remark that this is not a significant limitation on the flexibility of the method. In general, any splitting-style numerical method for computing probabilities of rare events will require some regime where the probabilities of events can be calculated directly, either by analytic methods or by Monte-Carlo estimation. All that is required is that a path of noise models is constructed that connects a target noise model to some `nice' noise model, such as the completely depolarizing channel, and that samples are collected from every noise model along the path.

\section{Additional numerical results \label{app:additional_numerical_results}} 

\begin{figure} 
	\includegraphics{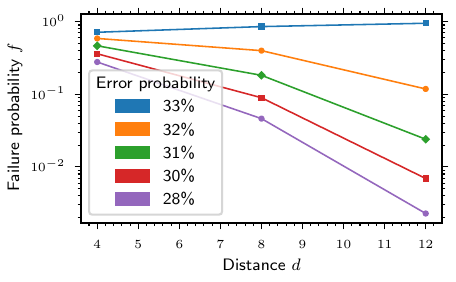} 
    \includegraphics{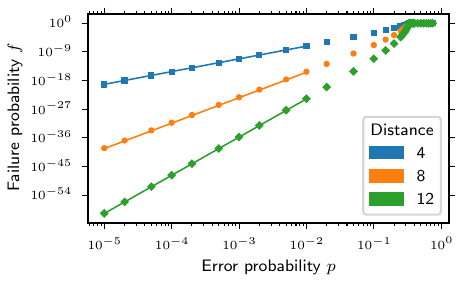} 
 
 \caption{\label{fig:f_zt_ft} Fault-tolerant logical failure probabilities for the zero-tolerance exclusive decoder. The decoder is used with the rotated surface code with $n = d^2$, and with $t = d$ rounds of syndrome measurement. The noise model is depolarizing noise of strength $p$, with measurement errors occuring with probability $p_m = 2 p / 3$. (top) We plot the logical failure probability against the code distance for a range of error probabilities near threshold, observing a threshold value of $p \approx 32\%$ (bottom) We show the scaling of logical failure probabilities below threshold. Markers correspond to data points, and the solid lines are the low-error probability ansatz Eq.~\eqref{eq:ztft_f}, along with Eq.~\eqref{eq:A(d)}.}
\end{figure} 

In this section, we expand on results relating to the zero-tolerance decoder. The additional results that we show are: 
\begin{enumerate} 
    \item We demonstrate fault-tolerant thresholds for logical failure probabilities using the zero-tolerance decoder, and show the subthreshold scaling behaviour of the failure probability. 
    \item We provide data on abort probabilities in the fault-tolerant regime. 
    \item We provide data on abort probabilities in the code-capacity regime, which was omitted from the main text on account of there being no thresholds for aborts.  
\end{enumerate} 

\subsection{Zero-tolerance fault-tolerant performance \label{app:fault_tolerant_comparative_decoder}} 

In this section we look at the fault-tolerant version of the exclusive decoder. We choose periodic time-like and rotated periodic space-like boundary conditions, which are often convenient for numerical work. We repeat $t = d$ rounds of syndrome measurement. The noise model is depolarizing noise of strength $p$, with measurement errors occuring with probability $p_m = 2 p / 3$. 

In Fig.~\ref{fig:f_zt_ft}, we demonstrate a fault-tolerant threshold value of $32\%$. This is smaller than the code-capacity threshold, which is to be expected as a general principle since there are more failure mechanisms in the fault-tolerant regime than there are in the code-capacity regime. 

\begin{figure} \includegraphics{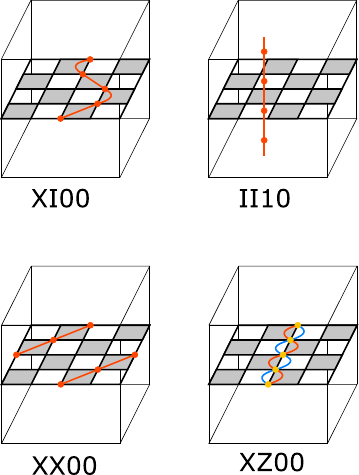}

    \vspace{0.75cm} 

    \begin{tabular}{|p{2cm}||p{2cm}|p{3cm}|  }
    \hline
    Sector&Multiplicity&Probability\\
    \hline
    $II00$ (x1) & $1$ & $(1 - p)^{d^2 t} (1 - p_m)^{d^2 t}$\\
    $XI00$ (x4) & $\frac{d t}{2} {\binom{d}{d/2}} $&$(\frac{p}{3})^d$\\
    $II10$ (x2) & $\frac{d^2}{2}$ & $(\frac{2p}{3})^d$\\
    $XX00$ (x2) & $d t $ & $(\frac{d}{3})^d$\\
    $XZ00$ (x2) & $d t$  & $ (\frac{p}{3})^d$\\
    \hline
    \end{tabular} 

    \vspace{0.5cm}

    \caption{\label{fig:ztft_errors} Analytic data used for the low-error probability ansatz Eq.~\eqref{eq:A(d)}, which sets the failure probability of the zero-tolerance decoder in the fault-tolerant regime as per Eq.~\eqref{eq:ztft_f}. Sectors are labelled first by a logical Pauli on each qubit, and then by the parity of Z-type (then X-type) measurements in any time-slice. (top) Example error configurations from each sector that contains a minimum-weight error. Red balls denote Pauli X-type errors or measurement faults on Z-type stabilizers. Yellow balls denote Pauli Y-type errors. (bottom) Tabulated data for each sector that contains a minimum-weight failing error. The times symbol next to each sector label counts the number of other sectors that are equivalent to the given sector by a combination of transversal hadamard and/or rotation.}
\end{figure} 

In the regime of small error probabilities, the logical failure probability is determined by the number of least-weight failing errors, and we propose the ansatz 
\begin{equation}
    f = A(d) p^d \label{eq:ztft_f} 
\end{equation} 

where $A(d)$ is related to the number of least-weight failing errors, and will be discussed momentarily. In Fig.~\ref{fig:ztft_abort}, we plot failure probabilities below threshold, and demonstrate good agreement with Eq.~\eqref{eq:ztft_f}. We observe that failures scale with $p^d$ instead of $p^{d/2}$, as would be the case with a standard decoder. In the fault-tolerant regime, just as in the code-capacity regime, 
the zero-tolerance decoder attains a quadratic improvement to logical failure probabilities below threshold, as compared to standard decoders. 

We now look more closely at the combinatoric pre-factor $A(d)$. We remark that the ansatz Eq.~\eqref{eq:ztft_f} is slightly more general than, for instance, Eq.~\eqref{eq:f}, since the form of $A(d)$ is as-yet unspecified. This more general ansatz is necessary, since an ansatz of form $A(d) = C A^d$ was not observed to satisfactorily fit the data for small code distances, and we require an accurate ansatz for fault-tolerant overhead estimation in Sec.~\ref{sec:overheads}. We now construct a sophisticated ansatz for $A(d)$ by explicitly counting least-weight failing errors. 

In the zero-tolerance regime, a logical failure must result from either a logical operator in the code, or a time-like string of measurement errors that wraps around the time-like direction. It is possible to count the number of least-weight failing errors in this setting exactly, and derive an ansatz for the post-selected failure rates that becomes exact in the low-error rate regime. For example, we consider the logical sector that involves no errors on the logical qubits, but with a non-trivial string of faulty $Z$-type measurement outcomes, as shown in the top right of Fig.~\ref{fig:ztft_errors}. These error chains must be straight in order to be least-weight. Therefore, there are only $d/2$ least-weight errors in this sector, one for each $Z$-type stabilizer in the code. Each of these least-weight errors has probability $p_m^d = (2/3)^d p^d$ of occurring, since they are made up of measurement errors. Therefore, failure mechanisms from this sector contribute a term $(d/2)(2/3)^d$ to the function $A(d)$. All the least-weight failing errors are tabulated in Fig.~\ref{fig:ztft_errors}, and a careful consideration of all the failure mechanisms in the code leads to the following ansatz

\begin{equation} 
    A(d) = \frac{4 d^2}{2} \binom{d}{d/2} \left(\frac{1}{3}\right)^d + d^2 \left(\frac{2}{3}\right)^d + 4 d^2 \left(\frac{1}{3} \right)^d \label{eq:A(d)}
\end{equation} 

We plot subthreshold failure probabilities in Fig.~\ref{fig:f_zt_ft}, and see good agreement with the ansatz for error probabilities $p \leq 10^{-3}$.

\subsection{Zero-tolerance fault-tolerant abort probabilities} 

\begin{figure} 
    \includegraphics{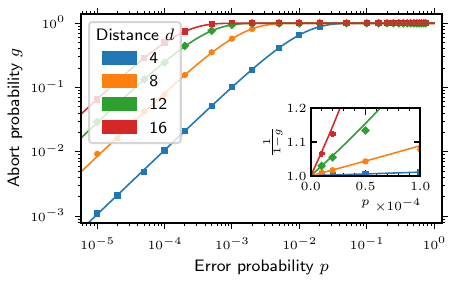} 
    \caption{\label{fig:ztft_abort} Fault-tolerant abort probabilities for the zero-tolerance decoder. The decoder is used with the rotated surface code with $n = d^2$ qubits, and with $t = d$ rounds of syndrome measurement. The noise model is depolarizing noise of strength $p$, with measurement errors occurring with probability $p_m = 2 p / 3$.
    Data points are markers, and the solid line is the ansatz Eq.~\eqref{eq:ztft_abort} . (inset) A zoomed in view of the inverse of the acceptance probability $(1 - g_0)^{-1}$. This is the factor increase in the time cost of the protocol as a result of exclusive decoding for one decoding cycle, since the computation may need to be repeated until the decoder accepts. 
    }
\end{figure} 

The fault-tolerant exclusive decoder at zero-tolerance does not possess a threshold for aborts, since increasing the code distance will always increase the probability that a single error will result in an aborted outcome. Despite the lack of threshold, the abort probability is still an important quantity that sets the cost of the scheme.

A first-order expression for the abort probability at zero-tolerance is given by the probability that any error occurs in the code. To first order, we have that the abort probability is given: 
\begin{equation} 
    g = 1 - (1 - p)^{t d^2} (1 - p_m)^{t d^2} \approx t d^2 (p + p_m) \label{eq:ztft_abort} 
\end{equation} 
where the approximation in the second equality holds when $p d^2 t \ll 1$. The abort probabilities for the zero-tolerance case are sufficiently high that they can be studied effectively using pure Monte Carlo sampling, and we compare Eq.~\eqref{eq:zero_tolerance_abort} to numerical data in Fig.~\ref{fig:ztft_abort}. We observe that in this heavily post-selected regime, the abort probability increases quadratically with the code distance at all error probabilities. 

\subsection{Zero-tolerance code-capacity abort probabilities} 

The zero-tolerance exclusive decoder in the code capacity setting does not possess a threshold for aborts. We remark that abort probabilities in the code-capacity setting for the zero-tolerance decoder are qualitatively identical to the abort probabilities in the fault-tolerant regime, which were discussed in the previous section. 

\begin{figure} 
	\includegraphics{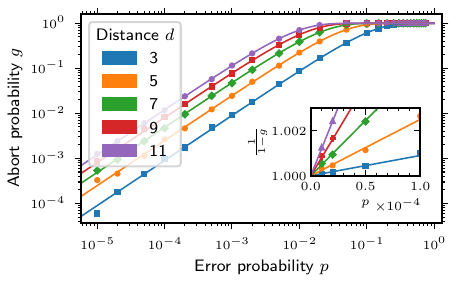} 
	\caption{\label{fig:ztcc_abort} Code-capacity abort probabilities for the zero-tolerance decoder. The decoder is used with the rotated surface code with $n = d^2$ qubits, and the noise model is depolarizing noise of strength $p$. Data points are markers, and the solid line is the ansatz Eq.~\eqref{eq:zero_tolerance_abort}. (inset) A zoomed in view of the inverse of the acceptance probability $(1 - g_0)^{-1}$. This sets the time-overhead of the exclusive decoder, since a computation may need to be repeated until the decoder accepts.} 
\end{figure}

A first order expression for the abort probability is given by considering the probability of any error occurring in the code:  

\begin{equation} 
    g = 1 - (1 - p)^{d^2} \approx p d^2 \label{eq:zero_tolerance_abort} 
\end{equation} 

where the approximation in the second equality holds when $p d^2 \ll 1$. We study the abort probabilities using Monte Carlo sampling, and compare Eq.~\eqref{eq:zero_tolerance_abort} in Fig.~\ref{fig:ztcc_abort}. As for the fault-tolerant case, there is no threshold behaviour and the abort probability increases quadratically with the code distance. The key difference between the fault-tolerant regime and the code-capacity regime is that abort probabilities are an order of magnitude smaller in the code-capacity regime, since there are fewer locations for errors to occur. 

\clearpage 

\section{Supporting figures \label{app:supporting_figures}} 

This section does not contain any additional results not explored in the main text. This section provides a more in-depth detailing of some of the data analysis and fitting procedures that support the main text figures. This section contains the following figures. 

\subsection{}

\begin{itemize} 
    \item Critical exponent plots of logical threshold values 
    \item Goodness of fit of exponential decay of failure probabilities below threshold
    \item Plot of offset $C(p)$ for failure probabilities below threshold. 
    \item Critical exponents plots of abort threshold values 
    \item Goodness of fit of exponential decay of aborts below threshold 
    \item Plot of offset $\tilde{C}(p)$ for abort probability below threshold 
    \item Goodness of fit of exponential decay of acceptance probability above the abort threshold. 
    \item Plot of offset $\tilde{C}_h(p)$ for acceptance probability above the abort threshold.  
\end{itemize} 

\begin{figure*} 
    \includegraphics{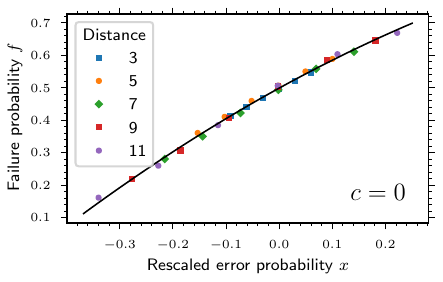}
    \includegraphics{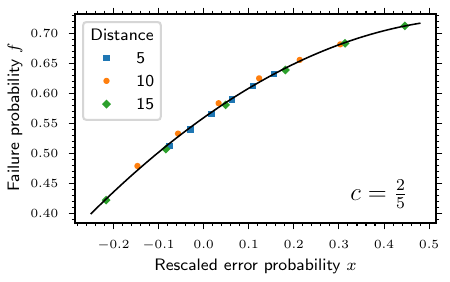}
    \includegraphics{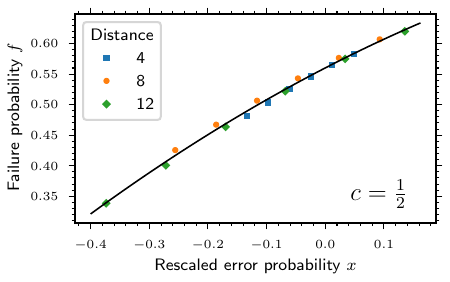}
    \includegraphics{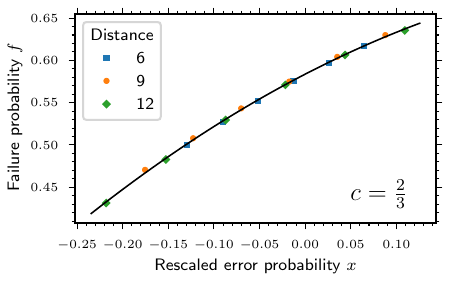}
    \includegraphics{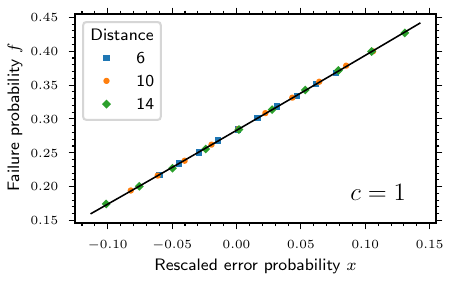}
    \caption{\label{fig:failure_threshold_extra}Critical exponent plots for threshold values of failure. For each tolerance, we fit an ansatz near threshold $f = Ax^2 + Bx + C$, where $x = (p - p_\text{th})d^{- \nu}$ is the re-scaled error probability, and $A, B, C, p_\text{th}, \nu$ are all extracted from fits. Markers are data points, plotted as a function of the re-scaled error probability, and black lines are the ansatz. The fitted parameter $p_\text{th}$ is the abort threshold that we report in the main text. }
\end{figure*} 

\begin{figure*} 
    \includegraphics{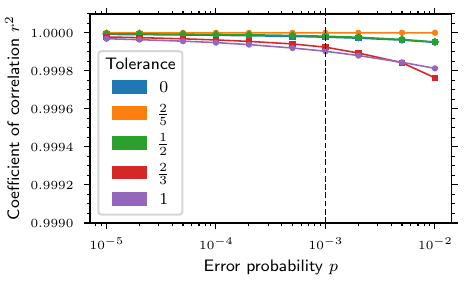}
    \includegraphics{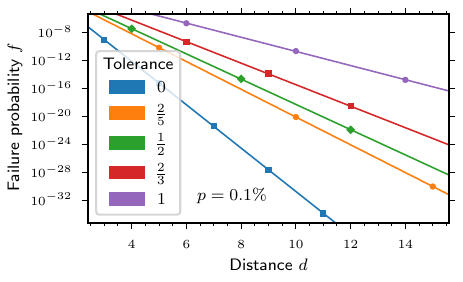}
    \caption{\label{fig:failure_extra12} Goodness of fit of a model describing an exponential decay of failure probabilities below the failure threshold, ie, the ansatz $\log f = d \log \Lambda(p) + \log C(p)$.  (top) For each error probability, we plot the coefficient of correlation resulting from a linear fit $\log f$ against $d$. (bottom) We plot the failure probability $f$ against the code distance $d$ at a subthreshold error probability $p = 10^{-3}$. Markers are data points and the solid lines are the ansatz $\log f = d \log \Lambda + \log C$, for $\Lambda$ and $C$ extracted by fitting. 
    }
\end{figure*} 

\begin{figure*} 
    \includegraphics{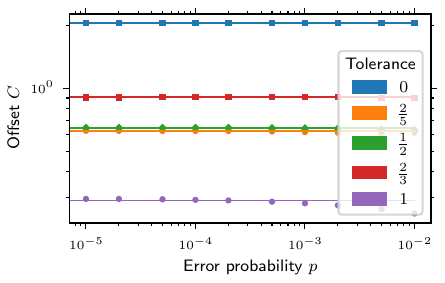} 
    \caption{\label{fig:failure_extra3} The offset function $C(p)$ is plotted, which is extracted from a linear fit of form $\log f = d \log \Lambda(p) + \log C(p)$, where $f$ is failure probabilities of the exclusive decoder, with the same code and error model as studied in the main text. Markers denote fitted values. We remark that low-error probability ansatz of Eq.~\eqref{eq:f} assumes that $C(p) \equiv C$ is a constant, and we plot this low-error probability ansatz using solid lines. 
    } 
\end{figure*} 

\begin{figure*} 
    \includegraphics{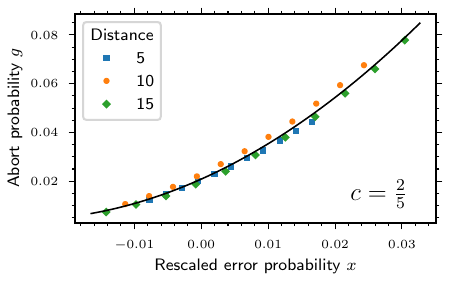}
    \includegraphics{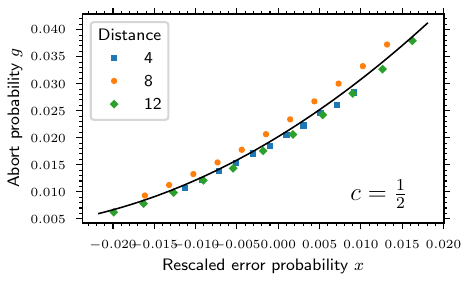} 
    \includegraphics{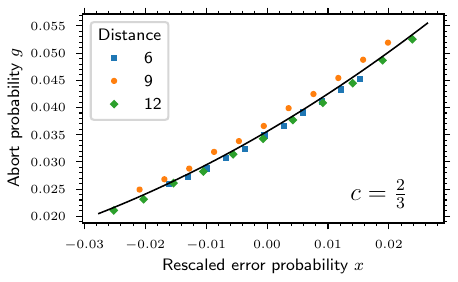} 
    \caption{\label{fig:abort_threshold_extra}Critical exponent plots for threshold values of abort. For each tolerance, we fit an ansatz near threshold $g = \tilde{A}x^2 + \tilde{B}x + \tilde{C}$, where $x = (p - \tilde{p}_\text{th})d^{- \tilde{\nu}}$ is the re-scaled error probability, and $\tilde{A}, \tilde{B}, \tilde{C}, \tilde{p}_\text{th}, \tilde{\nu}$ are all extracted from fits. Markers are data points, plotted as a function of the re-scaled error probability, and black lines are the ansatz. The fitted parameter $\tilde{p}_\text{th}$ is the abort threshold that we report in the main text. 
    }
\end{figure*} 

\begin{figure*} 
    \includegraphics{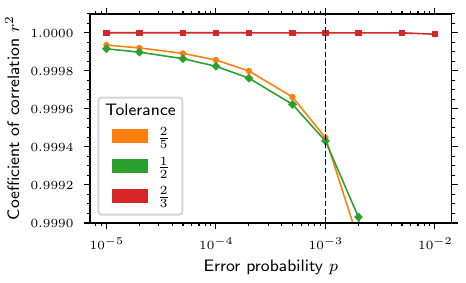} 
    \includegraphics{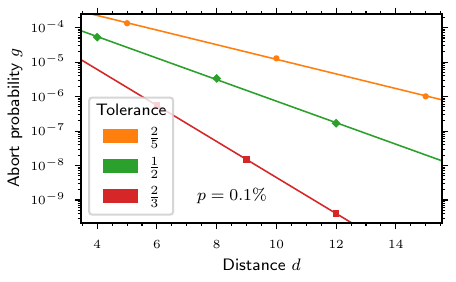} 
    \caption{\label{abort_subthreshold_extra} Goodness of fit of a model describing an exponential decay of abort probabilities below the abort threshold, ie, the ansatz $\log g = d \log \tilde{\Lambda}(p) + \log \tilde{C} (p)$.  (top) For each error probability, we plot the coefficient of correlation resulting from a linear fit $\log g$ against $d$. (bottom) We plot the failure probability $f$ against the code distance $d$ at a subthreshold error probability $p = 10^{-3}$. Markers are data points and the solid lines are the ansatz $\log g = d \log \tilde{\Lambda} + \log \tilde{C}$, for $\tilde{\Lambda}$ and $\tilde{C}$ extracted by fitting.}
\end{figure*} 

\begin{figure*} 
    \includegraphics{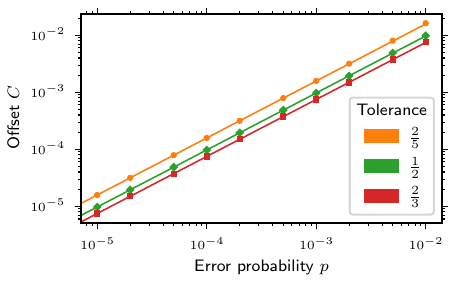} 
    \caption{\label{fig:abort_extra3} The offset function $\tilde{C}(p)$ is plotted, which is extracted from a linear fit of form $\log g = d \log \tilde{\Lambda}(p) + \log \tilde{C}(p)$, where $g$ is abort probabilities of the exclusive decoder, with the same code and error model as studied in the main text. Markers denote fitted values. We remark that low-error probability ansatz of Eq.~\eqref{eq:g} assumes that $\tilde{C}(p) = \tilde{C} p $ is proportional to the error probability, and we plot this low-error probability ansatz using solid lines. 
    } 
\end{figure*} 

\begin{figure*} 
    \includegraphics{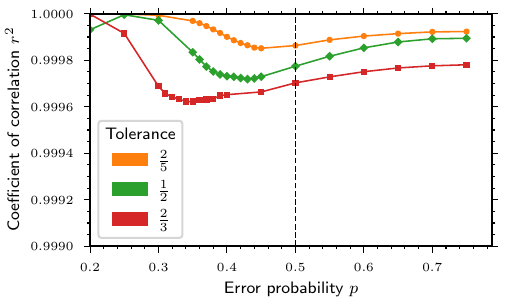} 
    \includegraphics{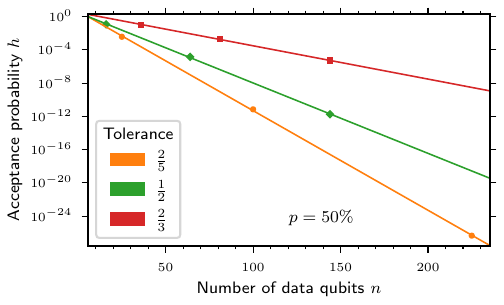} 
    \caption{\label{fig:abort_above_threshold_extra12} Goodness of fit of a model describing an exponential decay of acceptance probabilities above the abort threshold, ie, the ansatz $\log h = n \log \tilde{\Lambda}_h(p) + \log \tilde{C}_h (p)$, where $n$ is the number of data qubits with $n = d^2$ for the rotated surface code. (top) For each error probability, we plot the coefficient of correlation resulting from a linear fit $\log g$ against $n$. (bottom) We plot the acceptance probability $h$ against the number of qubits $n$ at an error probability $p = 50\%$. Markers are data points and the solid lines are the ansatz $\log g = n \log \tilde{\Lambda} + \log \tilde{C}$, for $\tilde{\Lambda}$ and $\tilde{C}$ extracted by fitting.}
\end{figure*} 

\begin{figure*} 
    \includegraphics{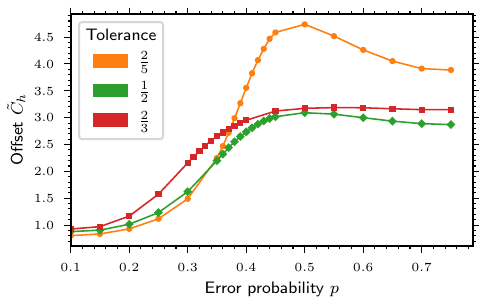} 
    \caption{\label{fig:abort_above_threshold_extra3}The offset function $\tilde{C}_h(p)$ is plotted, which is extracted from a linear fit of form $\log h = n \log \tilde{\Lambda})_h(p) + \log \tilde{C}_h(p)$, where $h$ is acceptance probabilities of the exclusive decoder, with the same code and error model as studied in the main text. Markers denote fitted values. } 
\end{figure*}

\clearpage

\section{Exclusive Union-Find \label{app:union-find}}

The notion of a exclusive decoder is not restricted to MWPM-based decoding, and is sufficiently general to be applied to a range of underlying standard decoders from QEC.  Here, we define a exclusive variant of the union-find (UF) decoder~\cite{delfosse2021almostlinear}. 

The first stage of the standard union-find algorithm is syndrome validation, whereby an initial error consisting of a Pauli part $E$ and erasure part $\sigma $ is mapped to a final erasure $\sigma'$. If the Pauli part $E$ is contained in the final erasure, and the final erasure does not include a logical operator, then a correction can be assured. 

To generalize this, we define the survived distance $d_\text{surv}$, which will measure the amount of uncertainty in a correction. The survived code distance is defined with reference to the final erasure $\sigma'$. If we take a logical operator $\bar{L}$, then it can be decomposed into a part with weight $w$ that shares no support with the final erasure, and a part with weight $w'$ that is entirely supported on the final erasure, so that $w + w' \geq d$. The surviving distance is then the minimum weight $w$ taken over all logical operators $\bar{L}$. We propose a decoder that aborts whenever 
\begin{equation} 
    1 - \frac{d_\text{surv}}{d} > c \label{eq:UF-abort}\,.
\end{equation} 

We show now that this decoder can either successfully correct, or abort, all errors consisting of a weight $s$ Pauli part $E$, and a weight $t$ erasure part $\sigma$, provided $(2s + t)/2 < kd$, with $k$ given 
\begin{equation} 
    k = 1 - \frac{c}{2} \,.
\end{equation} 

The UF decoder breaks the initial erasure $\sigma$ up into connected components, called clusters, and iteratively updates each cluster. Let $l$ denote the sum of the diameters of the clusters at some point in the algorithm, and let $w'$ denote the weight of $E$ supported on the clusters at that point. The growth of a cluster is always accompanied by the discovery of at least one new qubit in the support of $E$, and its inclusion into the cluster. This leads to a lower bound $w' \geq (l - t) / 2 $. Further, if $(E, \sigma)$ results in failure, then we must also have the lower bound $w \geq d_\text{surv}$, where $w$ is the weight of $E$ not supported on the final clusters. Writing $s = w + w'$, we can write 
\begin{equation} 
    2s + t \geq 2d_\text{surv} + l \geq d_\text{surv} + l \geq 2d\left( 1 - \frac{c}{2} \right) \,.
\end{equation}  
The first inequality above is the application of the bounds discussed in the previous paragraph, which assume that the result of the UF algorithm is a non-trivial logical operator. The second inequality uses the bound $d_\text{surv} + l \geq d$, which is true since the minimization in the definition of $d_\text{surv}$ can be carried out without loss of generality by considering only logicals that do not enter or exit each final cluster more than once. The last inequality follows from the assumption that the decoder does not abort, and is the negation of Eq.~\eqref{eq:UF-abort}.

\end{document}